\input{aipcheck}

\documentclass[
    ,final            
  ]
  {aipproc}

\layoutstyle{6x9}

\newcommand{\fr}[1]{
             \frac{#1}}
\newcommand{\bea}{\begin{eqnarray}}
\newcommand{\eea}{\end{eqnarray}}

\newcommand{\chibar}{\overline{\chi}}

\newcommand{\ket}{{\rangle }}

\newcommand{\bra}{{\langle }}
\newcommand{\gc}{\bra\fr{\alpha_s}{\pi}G^2\ket}

\newcommand{\qc}{\bra\,\overline{q}q\,\ket}

\newcommand{\ga}{{g_{{\mathcal A}}}}

\begin{document}

\title{ Non-factorizable contributions to 
  $\overline{B^0_d} \rightarrow D_s^{(*)} \overline{D_s^{(*)}}$  
from chiral loops and tree level $1/N_c$ terms\footnote{Presented by
J.O. Eeg}}

\classification{13.20.Hw, 12.39-St, 12.39.Fe,12.39.Hg}
\keywords      {B-decays, chiral perturbation theory}

\author{J.O. Eeg}{
  address={Dept. of Physics, Univ. of Oslo, P.O. Box 1048
Blindern, N-0316 Oslo, Norway}
}

\author{S. Fajfer ,}{
  address={Department of Physics, Univ. of Ljubljana, Jadranska 19,
 and \\
J. Stefan Institute, Jamova 39, P.O. Box 3000, 1001 Ljubljana, Slovenia}
}

\author{A. Prapotnik Brdnik}{
  address={Faculty of Civil engineering, Univ. of Maribor,
  Smetanova ul. 17, 2000 Maribor, Slovenia} }

\begin{abstract}

We point out that the amplitudes for the decays  
$\overline{B}^0 \to D_s^+  D_s^-$ and $\overline{B}^0_s \to D^+  D^-$
have no factorizable contributions.
If one or two of the $D$-mesons in the final state are vectors (i.e $D^*$ 's)
there are  relatively small factorizable contributions through the
annihilation mechanism.
The dominant contributions to the decay amplitudes 
arise from  chiral loop contributions and 
$1/N_c$ suppressed tree level.
We predict that the 
branching ratios   for the processes $\bar B^0_d \to D_s^+ D_s^-$, 
  $\bar B^0_d \to D_s^{+*} D_s^- $ and 
 $\bar B^0_d \to D_s^+ D_s^{-*}$ are all of order $(2- 3) \times 10^{-4}$,
while $\bar B^0_s \to D_d^+ D_d^-$,
  $\bar B^0_s \to D_d^{+*} D_d^- $ and 
 $\bar B^0_s \to D_d^+ D_d^{-*}$ are  of order $(4- 7) \times 10^{-3}$.
 If {\em both}  $D$-mesons  in the final state
  are   $D^*$'s,  we obtain branching ratios of order
two times bigger.

\end{abstract}

\maketitle

\section{Introduction}

Decay modes like $B \rightarrow \pi \pi, K \pi$ are intensively
studied. From the theoretical side, for instance within
QCD-factorization.  
The decays of a $B$- meson into two $D$-mesons are different because the
energy release is only of order 1 GeV, and
therefore (QCD-) factorization is not expected to hold. $B \to DD$ decays
are also studied  experimentally \cite{exp}.
Here we discuss non-factorizable contributions to 
 the decay modes  $\overline{B^0_d} \rightarrow D_s^{(*)}
 \overline{D_s^{(*)}}$, where $ D_s^{(*)}$ is a pseudoscalar or a
 vector meson.
At quark level
such decays occur through the 
annihilation mechanism
$b \bar{q} \rightarrow c \bar{c}$, where $q=d,s$ 
respectively. However, in the factorized limit the annihilation
 mechanism will give a vanishing amplitude 
 due to current conservation (similar to $D^0 \rightarrow K^0
 \overline{K^0}$ \cite{EFZ}), unless one or two of the $D$-mesons in
 the final state are vectors. The contributions due to the 
annihilation mechanism 
 are proportional to a numerically non-favored  Wilson coefficient.
In contrast, 
 the typical   factorized   decay modes  which proceed
through the spectator mechanism, 
$\overline{B^0} \rightarrow D^+ D_s^-$ say, 
are proportional to the numerically favored   Wilson 
coefficient.

In our approach  \cite{EFH,EFP}, the non-factorizable  
contributions are coming from  
chiral loops and from tree level amplitudes generated by soft gluon
 emision forming a  gluon condensate.
The  gluon condensate contributions
can be 
calculated within 
a recently developed Heavy Light Chiral Quark Model (HL$\chi$QM) \cite{ahjoe}.
This model has been applied to processes involving $B$-mesons in 
\cite{ahjoeB,Bet}.
Both the chiral loop contributions and the gluon condensate
 contributions are 
 proportional to the numerically favorable Wilson coefficient.

\section{FRAMEWORK}

\subsection{Effective Lagrangian at quark level}

The relevant effective Lagrangian at quark level reads:
\begin{equation} {\mathcal L}_{W}=  - \frac{G_F}{\sqrt{2}} V_{cb}V_{cq}^*
\sum_i a_i(\mu) \; Q_i (\mu) \,,
\label{Lquark}
\end{equation}
where $q=d,s$ and $a_i(\mu)$ are  Wilson coefficients that
carry all information of the short distance  physics above the
renormalization scale $\mu$.
 The matrix elements of $Q_i(\mu)$ 
contain all non-perturbative, long distance  physics below
$\mu$.
Within Heavy Quark Effective Theory~(HQEFT)
the effective non-leptonic 
Lagrangian ${\mathcal L}_{W}$ can be  evolved down to the scale
 $\mu = \Lambda_\chi \simeq$ 1 GeV \cite{GKMWF} .

 The  numerically relevant operators in our case are 
\begin{eqnarray}
Q_{1}= 4(\overline{q}_L \gamma^\mu b_L) \; ( \overline{c}_L
\gamma_\mu c_L ) \quad , \quad
Q_{2}= 4( \overline{c}_L \gamma^\mu b_L ) \; ( \overline{q}_L
\gamma_\mu c_L ) \,, 
\label{Q12} 
\end{eqnarray}  
where  $L$
denotes a left-handed particle.
At $\mu =
\Lambda_\chi$,  which by construction is the matching
scale within our approach \cite{EFH,ahjoe,ahjoeB}, 
one finds $a_1 \simeq -0.35 -0.07i$
and $a_2 \simeq 1.29+ 0.08i$. Note that  the Wilson
coefficients $a_i$ are complex  below $\mu=m_c$ \cite{GKMWF}. 
 In the next subsection we will see how
 the currents in the operators in (\ref{Q12}) are bosonized.

In order to obtain all matrix elements of
the Lagrangian (\ref{Lquark}) we need the 
 Fierz transformed version of the operators in  (\ref{Q12}).
To find  these, we use the relation:
\begin{equation}
\delta_{i j}\delta_{l n}  =   \frac{1}{N_c} \delta_{i n} \delta_{l j}
 \; +  \; 2 \; t_{i n}^a \; t_{l j}^a \, ,
\label{fierz}
\end{equation}
where $i$, $j$, $l$ and $n$ are color indices running from 1 to 3 and
 $t^a$ denotes the color matrices, $a$ being the color octet index. One obtains
\begin{equation}
Q_1^F = \frac{1}{N_C} Q_2 + \widetilde Q_2 \; ,
  \quad Q_2^F = \frac{1}{N_C} Q_1 + \widetilde Q_1 \; ,
\label{Q12F}
\end{equation}
where the superscript $F$ means ``Fierzed'', and 
\begin{eqnarray}
\widetilde{Q_{1}}  = 4  (\overline{q}_L \gamma^\mu t^a  b_L )  \; \,
           ( \overline{c}_L \gamma_\mu t^a c_L )  \quad , \quad
\widetilde{Q_{2}}  =  4 \,  ( \overline{c}_L \gamma^\mu t^a b_L )  \; \,
           ( \overline{q}_L \gamma_\mu t^a c_L ) \, .
\label{QCol} 
\end{eqnarray}  
 These operators generate contributions proportional to the 
 gluon condensate.

\subsection{Heavy Light Chiral Perturbation Theory}

The Heavy Quark Effective Theory (HQEFT) Lagrangian is:
\begin{equation}
{\mathcal L}_{HQEFT} = \pm \overline{Q_v^{(\pm)}} \, i v \cdot D \,
Q_v^{(\pm)} + {\mathcal O}(m_Q^{- 1}) \; ,
\label{LHQEFT}
\end{equation}
where 
 $Q_v^{(+)}(x)$ is a (reduced)  heavy quark field ($b$ or $c$ in our
 case) with velocity $v$, and  $Q_v^{(-)}(x)$ is the field of a heavy
 anti-quark ($\bar{c}$ in our case). Furthermore, $m_Q$ is the heavy
 quark mass, and
 $D_\mu$ is the covariant derivative containing the gluon field.

After integrating out the heavy and light quarks, the effective Lagrangian
up to ${\mathcal O}(m_Q^{-1})$ can be written as a kinetic term plus a
term
describing the chiral interaction between heavy and light mesons
\cite{ahjoe,itchpt}:
\begin{eqnarray}
{\mathcal L}_\chi  = - \ga \, Tr\left[\overline{H^{(\pm)}_{a}}H^{(\pm)}_{b}
\gamma_\mu\gamma_5 {\mathcal A}^\mu_{ba}\right]\, 
,\label{LS1}
\end{eqnarray}
where $H_a^{(\pm)}$ is the heavy meson field containing a spin zero
and a spin one boson:
\begin{eqnarray}
 H_a^{(\pm)}   =   P_{\pm} (P_{a \mu}^{(\pm)} \gamma^\mu - i P_{a
5}^{(\pm)} \gamma_5) \; \; .
\label{barH}
\end{eqnarray}
Here $a,b$ are
flavor indices and  $P_\pm=(1 \pm \gamma \cdot v)/2$ are projecting operators.  
 The  axial vector field  ${\mathcal A}_{\mu}$ in (\ref{LS1}) is defined as:
\begin{equation}
 {\mathcal A}_\mu =  -
\frac{i}{2} (\xi^\dagger\partial_\mu\xi -\xi\partial_\mu\xi^\dagger)\,, 
\label{defVA}
\end{equation}
where $ \xi\equiv exp[i(\Pi/f)]$. Moreover,  $f$ is the bare pion
 coupling
 and $\Pi$ is a 3 by 3 matrix
which contains the Goldstone bosons $\pi,K,\eta$ in the standard way,
 and  $\ga$ is the axial chiral coupling.

\begin{figure}[t]
\includegraphics[width=6cm]{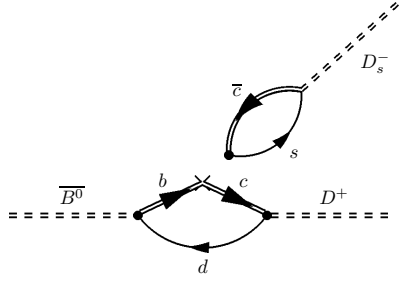}
\caption{\small{Factorized contribution for 
$\overline{B^0}  \rightarrow D^+ D_s^-$
through the spectator mechanism, which does not exist for
 decay mode $\overline{B^0} \rightarrow D_s^+ D_s^-$. There are
 similar diagrams with vector mesons.}}
\label{fig:bdd_fact}
\end{figure}

\begin{figure}[t]
\includegraphics[width=6cm]{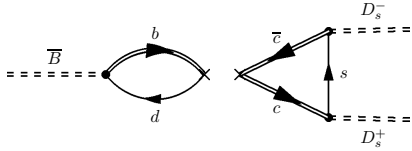} 
\caption{\small{Factorized contribution for 
$\overline{B^0}  \rightarrow D_s^+ D_s^-$
through the annihilation  mechanism, which give zero contributions if
both $D_s^+$ and $D_s^-$ are pseudoscalars.}}
 \label{fig:bdd_fact2}
\end{figure}

Based on the symmetry of HQEFT, we  obtain the bosonized currents. 
For a decay of the $b \bar{q}$ system (see Fig.~\ref{fig:bdd_fact2}, left)
we have \cite{ahjoe}:
\begin{equation}
 \overline{q_L} \,\gamma^\mu\, Q_{v_b}^{(+)} \; \longrightarrow \;
 \frac{\alpha_H}{2} Tr\left[\xi^{\dagger} \gamma^\alpha L \, H_{b}^{(+)}
 \right] \; ,
\label{J(0)}
\end{equation}
where (up to QCD and $1/m_Q$ corrections\cite{ahjoe})
 $\alpha_H=f_H \sqrt{m_H}$ for $H=B,D$. Further,
 $Q_{v_b}^{(+)}$ is the heavy $b$-quark field, $v_b$ is
its velocity, and $H_{b}^{(+)}$ is the corresponding heavy meson
field.
For the  $W$-boson materializing to a $\overline{D}$, the bosonized current 
$ \overline{q_L} \gamma^\mu\  Q_{{v_{\bar c}}}^{(-)}$ is  also given by 
 (\ref{J(0)}),  but with  $H_{b}^{(+)}$ replaced by 
$H_{\bar{c}}^{(-)}$ representing  the  $\overline{D}$ meson.
$Q_{{v_{\bar c}}}^{(-)}$ is the field of the heavy $\bar{c}$ quark,
 and $v_{\bar c}$ is its velocity
 (see Fig.~\ref{fig:bdd_fact}, right).

The  bosonized $b \rightarrow c$ transition current in
Fig.~\ref{fig:bdd_fact} is given by
\begin{equation}
 \overline{Q_{v_b}^{(+)}} \,\gamma^\mu  L Q_{v_c}^{(+)}\;\longrightarrow
 \; - \zeta(\omega) Tr\left[ \overline{H_c^{(+)}}\gamma^\alpha L
 H_{b}^{(+)} \right] , 
\label{Jbc}
\end{equation}
where $\zeta(\omega)$ is the Isgur-Wise function for the $\bar{B}
\rightarrow D$ - transition, and $v_c$ is the velocity of the heavy
$c$-quark.  Furthermore, $\omega \equiv v_b \cdot v_c= v_b \cdot
v_{\bar c} = M_B/(2M_D)$.  
For the weak current for $D \overline{D}$ production (corresponding to the
factorizable annihilation mechanism in Fig.~\ref{fig:bdd_fact2}), the current 
$ \overline{Q_{v_c}^{(+)}} \,\gamma^\mu\, L Q_{v_{\bar c}}^{(-)}$
is given by (\ref{Jbc}) with $H_{b}^{(+)}$ replaced by 
$H_{\bar{c}}^{(-)}$, and 
$\zeta(\omega)$ is replaced by $\zeta(-\lambda)$,
where $\lambda= v_{\bar c} \cdot v_c = [M_B^2/(2M_D^2) -1]$.
Note that $\zeta(-\lambda)$ is a complex function which is less known
than $\zeta(\omega)$.

The factorized contributions for the spectator and annihilation
diagrams are shown in the Figs.~\ref{fig:bdd_fact}  and  \ref{fig:bdd_fact2}.
The first diagram does not give any (direct) contributions to the class of processes
we consider, but is still important because it is the basis of our
chiral loops. 

\begin{figure}[t]
\includegraphics[width=7.5cm]{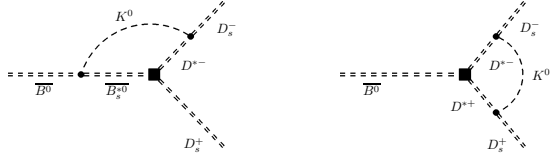}
\caption{Non-factorizable chiral loops for 
$\overline{B^0} \rightarrow D_s^+ D_s^-$. There are similar diagrams
  for vector mesons in the final state. The two diagrams illustrate
  the two classes of diagrams.}
\label{fig:chiral1}
\end{figure}

The chiral loop amplitudes  visualized in Fig.~\ref{fig:chiral1} are
 of order $(\ga m_K/4 \pi f)^2$ compared  to  typical factorizable 
amplitudes in processes where these exist.
For instance , the ratio between the chiral loop amplitude for
$\overline{B^0} \rightarrow D_s^+ D_s^-$ and the factorized amplitude
 for   $\overline{B^0} \rightarrow D^+ D_s^-$
 is $\simeq \; - 0.20 + 0.26 i \,$ (before the difference 
in KM structure is taken into account).
For vectors (i.e. $D^*$ 's) in the final state there are similar
diagrams with various combinations of pseudoscalars and vectors in the
loop, but the diagrams in Fig.~\ref{fig:chiral1} constitute the two
classes of diagrams \cite{EFP}.

 In a complete analysis, counterterms to the chiral loops has to be included.
These counterterms are not considered here (or in \cite{EFH,EFP}) and
has to be be considered together with the constant (non-logarithmic)
chiral loop terms which we also have dropped in this analysis. 
The inclusion of counterterms and constant chiral loops terms
 will be discussed elsewhere.

One might also write down possible  terms consistent with HQEFT
and chiral symmetry, for instance the following three terms:
 \bea
Tr\left[\xi^{\dagger} \sigma^{\mu \alpha}
L \,  H_{b}^{(+)} \right]  \cdot 
 Tr\left[
 \overline{H_c^{(+)}} \gamma_\alpha L  H_{\bar{c}}^{(-)} \gamma_\mu \right]
\quad , \quad
Tr\left[\xi^{\dagger}
L \,  H_{b}^{(+)} \right] \cdot
 Tr\left[
 \overline{H_c^{(+)}} \gamma^\alpha L  H_{\bar{c}}^{(-)} \gamma_\alpha
 \right] \; ,
\nonumber \\
\varepsilon^{\mu \nu \alpha \lambda} (v_c+\bar{v})_\nu 
Tr\left[\xi^{\dagger} \gamma^\mu
L \,  H_{b}^{(+)} \right] \cdot
 Tr\left[
 \overline{H_c^{(+)}} \gamma^\alpha L  H_{\bar{c}}^{(-)} \gamma_\lambda \right]
\; . \hspace{3cm}
\label{oneoverN}
\eea
Such terms do not appear in the factorized limit, and will correspond to
(at least) $1/N_c$ suppressed terms. Within pure chiral perturbation theory their
coefficients are unknown, but they might be calculated within  the
HL$\chi$QM, as described in the next subsection.

\subsection{The Heavy Light Chiral Quark Model}

The HL$\chi$QM Lagrangian is 
\begin{equation}
{\mathcal L}_{\rm{HL\chi QM}} = {\mathcal L}_{HQEFT} +
{\mathcal L}_{\chi QM} + {\mathcal L}_{Int} \; .
\label{totlag}
\end{equation}
The first term is given in (\ref{LHQEFT}) and the second term 
 is described by the Chiral Quark Model
of the light sector \cite{BEF} involving 
interactions between quarks and (Goldstone) mesons:
\begin{equation}
{\mathcal L}_{\chi QM} =
\bar \chi \left[\gamma^\mu (i {\cal D}_\mu +   
\gamma_5  {\mathcal A}_{\mu}) - m \right]\chi  \; .
\label{chqmR}
\end{equation}
Here $m=(230\pm20)$MeV is the $SU(3)$ invariant constituent light
quark mass, and $\chi$ is the flavor rotated quark field given by
$\chi_L=\xi^\dagger q_L$ and $\chi_R = \xi q_R$, where $q^T = (u,d,s)$
is the light quark field. 
The covariant derivative ${\cal D}_\mu$  contains the soft
gluon field forming the gluon condensates (besides some chiral interactions) 
\cite{ahjoe,ahjoeB,BEF}. 

The interaction between heavy meson fields and quarks is
described by \cite{ahjoe}:
\begin{equation}
{\mathcal L}_{Int}  =   
 -   G_H  \left[ \chibar_a \, \overline{H_a^{(\pm)}} 
\, Q^{(\pm)}_{v} \,
  +     \overline{Q_{v}^{(\pm)}} \, H_a^{(\pm)} \, \chi_a \right] ,  
\label{Int}
\end{equation}
where the coupling constant $G_H = \sqrt{2 m
\rho}/f$, 
 and $\rho$ is a hadronic parameter
depending on $m$ (numerically  $\rho$ is of order one). For further
details, see ref. \cite{ahjoe}.

\begin{figure}[t]
\includegraphics[width=6cm]{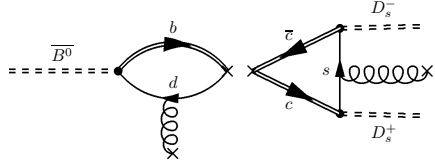}
\caption{Non-factorizable contribution for 
$\overline{B^0}  \rightarrow D_s^+ D_s^-$
through the annihilation mechanism with additional soft gluon emision.
 The wavy lines represent soft
gluons ending in vacuum to make gluon condensates.}
\label{fig:bdd_nfact2}
\end{figure}

The gluon condensate amplitudes can be written, within the
framework presented in the previous section, in a quasi-factorized
way as a product of matrix elements of colored currents in (\ref{QCol}),
 as visualized
in Fig.~\ref{fig:bdd_nfact2}.  The left part of
Fig.~\ref{fig:bdd_nfact2} corresponds to the bosonized colored current:
\bea
\left(\overline{q_L}\, t^a  \,\gamma^\alpha \, Q_{v_b}^{(+)}\right)_{1G} 
\;   \longrightarrow \; 
- \fr{G_H \, g_s}{64 \pi} \,G_{\mu\nu}^a \nonumber \\
\times Tr\left[\xi^\dagger
\gamma^\alpha L \, H_b^{(+)}
\left( \sigma^{\mu\nu} \, - \, F \,  \{\sigma^{\mu\nu},
 \gamma \cdot v_b \} \, \right)\right] , 
\label{1G}
\eea
where $G^a_{\mu \nu}$ is the octet gluon tensor, and
$F \; \equiv \; 2 \pi f^2/(m^2\,N_c)$
is a dimensionless quantity of the order 1/3. The symbol $\{\; , \; \}$
denotes the anti-commutator.
For the creation of a $D \overline{D}$ pair in the right part of Fig.
\ref{fig:bdd_nfact2},  the colored current 
$ \left(\overline{Q_{v_c}^{(+)}} \,t^a \; 
\gamma^\alpha \, L Q_{v_{\bar{c}}}^{(-)}\right)_{1G} \;$
is bosonized similarly to (\ref{1G}), but involves 
$H_{c}^{(+)}$ and $H_{\bar{c}}^{(-)}$.

Multiplying the two colored  currents, and introducing  gluon condensate contributions
by  the replacement:
\begin{equation}
g_s^2 G_{\mu \nu}^a G_{\alpha \beta}^a  \; \rightarrow 4 \pi^2
 \gc \frac{1}{12} (g_{\mu \alpha} g_{\nu \beta} -  
g_{\mu \beta} g_{\nu \alpha} ) \; , 
\label{gluecond}
\end{equation}
we obtain a bosonized effective Lagrangian which is $1/N_c$
suppressed compared to the factorized contributions. 
This effective Lagrangian corresponds to a certain linear combination
of  a priori possible  $1/N_c$ suppressed terms   at tree level (in
the chiral perturbation theory sense). Among these are the three terms
in (\ref{oneoverN}).

$1/m_b$ suppressed terms seem to be negligible.
In order to include $1/m_c$ terms, one must for instance consider the
 $c \bar{c}$ production current to this order:
\bea
\Delta J_\mu(\bar{c} c) \, = \, 
 \frac{1}{m_c} \overline{Q_{v_c}^{(+)}} \left( 
i \gamma \cdot \overleftarrow{D}_\perp(v_c) \,\gamma^\mu\, L
\, + \, \gamma^\mu\, L \, i \gamma \cdot D_\perp(v_{\bar{c}}) \right)
Q_{v_{\bar{c}}}^{(-)}
\; ,
\eea
where $D_\perp^\mu(v) = (g^{\mu \nu} - v^\mu v^\nu) D_\nu$.
Within the HL$\chi$QM one may estimate both factorizable and non-factorizable
$1/m_c$ corrections due to this current 
There are also other operators. Compared to other contributions studied here,
the relative size  of $1/m_c$ suppressed contributions 
will be of order $\tilde{m}/m_c$, where $\tilde{m}$
is some hadronic parameter within the HL$\chi$QM with dimension mass, such as
linear commbinations of  $m$ and  $\qc/f^2$. The total contributions from such terms are
 significant and will be studied elsewhere.

\section{Results and discussion}

In our calculation we used the following input parameters: 
$\alpha_B \simeq \alpha_D \simeq 0.33$ GeV$^{-3/2}$, $G_H=7.5$ GeV$^{-1/2}$ 
and $\gc= [(315\pm20)$ MeV]$^{4}$
\cite{ahjoe,ahjoeB}, $\ga=0.6$, $f_\pi=93$ MeV. 
 We find \cite{EFP} the following branching ratios
\begin{eqnarray}
Br(\bar B^0 \to D_s^+ D_s^-)=2.5 \times 10^{-4}\,,
 \qquad Br(\bar B^0_s \to D^+ D^-)=4.5 \times 10^{-3}\,,\\
Br(\bar B^0 \to D_s^{+*} D_s^-)=3.3 \times 10^{-4}\,,\qquad 
 Br(\bar B^0_s \to D^{+*} D^-)=6.8 \times 10^{-3}\,,\\
Br(\bar B^0 \to D_s^+ D_s^{-*})=2.0 \times 10^{-4}\,,\qquad 
Br(\bar B^0_s \to D^+ D^{-*})=4.3 \times 10^{-3}\,, \\
Br(\bar B^0 \to D_s^{*+} D_s^{-*})=5.4\times 10^{-4} 
\,,\qquad Br(\bar B^0_s \to D^{*+} D^{-*})=9.1 \times 10^{-3}\,.
\end{eqnarray}

The contribution of the constant term and the corresponding counterterm
can change the branchig ratio for $B$-meson decaying into two pseudoscalars by about
$10\%$, while in the case of decay  into one pseudoscalar and one
vector $D$-meson, this contribution is  in the range of  
$20-40\%$. 
In the case of $B$-meson decaying  into two vector mesons, the 
constant term is
estimated to be 2-8 times larger than the logaritmic contribution,
depending on  the  choice of the scheme in which the products of two Levi-Civita 
terms 
are considered.
The uncertainty in input parameters can result in an additional error 
 for the branching ratios. We estimate that it can be of the order of 
$~ 20\%$. 
Within our approach the 
 $1/m_Q$ corrections, with  $Q=c,b$   have  so far been omitted. 
At least the $1/m_c$ corrections will be numerically significant.

\begin{theacknowledgments}
 J.O.~Eeg is supported in part by the Norwegian
 research council
 and  by the European Union RTN
network, Contract No. HPRN-CT-2002-00311  (EURIDICE). 
S.~Fajfer is supported in part
 by the Ministry of Education,
Science and Sport of the Republic of Slovenia.

\end{theacknowledgments}


\begin{thebibliography}{9}

\bibitem{exp}

For a recent study, see K. Abe et al.,
e-Print Archive: hep-ex/0508040.

\bibitem{EFZ}
J. O. Eeg, S. Fajfer, J. Zupan,{\it  Phys. Rev.}  {\bf D 64}, 034010 (2001).  

\bibitem{EFH}
J.O. Eeg, S. Fajfer, A. Hiorth, {\it Phys. Lett. } {\bf B 570} (2003)  46-52.  

\bibitem{EFP}
J.O. Eeg, S. Fajfer, A. Prapotnik, 
{\it  Eur.Phys.J.} {\bf C42} (2005) 29-36.

\bibitem{ahjoe}
A.~Hiorth and J.~O.~Eeg, {\it Phys. Rev.} {\bf D 66} 
(2002) 074001, and references therein. 


\bibitem{ahjoeB}
 A.~Hiorth and J.~O.~Eeg, 
{\it Eur.Phys.J.direct} {\bf C30} (2003) 006.

\bibitem{Bet}
J.O.~Eeg,  A.~Hiorth,and  A.D.~Polosa
{\it Phys.Rev.} {\bf D65} (2002) 054030.


J.O.~Eeg, K.~Kumericki, and  I.~Picek, 
e-Print Archive: hep-ph/0506152.


\bibitem{GKMWF}
B.~Grinstein, W.~Kilian, T.~Mannel, and M.B.~Wise, {\it Nucl. Phys.} {\bf B363}
(1991) 19.

R.~Fleischer, {\it Nucl. Phys.} {\bf B 412} (1994) 201.

\bibitem{itchpt}
 R.~Casalbuoni, A.~Deandrea, N.~Di~Bartelomeo, 
R.~Gatto, F.~Feruglio and  G.~Nardulli, \\
{\it Phys. Rep.} {\bf 281},  145 (1997);



\bibitem{BEF}
See for example: 
A.~Pich and E.~de Rafael, {\it Nucl. Phys}.   
{\bf B 58}, 311 (1991), 
S.~Bertolini, J.O.~Eeg and M.~Fabbrichesi, 
{\it Nucl. Phys.} {\bf B449} (1995) 197, 
V.~Antonelli, S.~Bertolini, J.O.~Eeg,
M.~Fabbrichesi and E.I.~Lashin, 
 {\it Nucl. Phys.} {\bf B469} (1996) 143,
S.~Bertolini, J.O.~Eeg, M.~Fabbrichesi, and  E.I.~Lashin,
{\it Nucl.Phys.} {\bf B514} (1998) 63-92,
 and references therein. 


\end{thebibliography}
\end{document}